# Predictability in an 'Unpredictable' Artificial Cultural Market

## Paul Ormerod[*] and Kristin Glass[**]

### August 2009

**\*** corresponding author, Volterra Consulting, London, UK, pormerod@volterra.co.uk
\*\* New Mexico Institute of Mining and Technology, New Mexico, USA
kglass@icasa.nmt.edu

*Abstract*

*In social, economic and cultural situations in which the decisions of individuals are influenced directly by the decisions of others, there appears to be an inherently high level of* ex ante *unpredictability. In cultural markets such as films, songs and books, well-informed experts routinely make predictions which turn out to be incorrect.*

*We examine the extent to which the existence of social influence may, somewhat paradoxically, increase the extent to which winners can be identified at a very early stage in the process. Once the process of choice has begun, only a very small number of decisions may be necessary to give a reasonable prospect of being able to identify the eventual winner.*

*We illustrate this by an analysis of the music download experiments of Salganik et.al. (2006). We derive a rule for early identification of the eventual winner. Although not perfect, it gives considerable practical success. We validate the rule by applying it to similar data not used in the process of constructing the rule.*

1.   **Introduction**

Enormous resources are devoted to the task of predicting the outcome of social processes in domains such as economics, public policy, and popular culture. But these predictions are often woefully inaccurate. Consider, for instance, the case of cultural markets. Perhaps the two most striking characteristics of cultural markets, for example, are their simultaneous *inequality*, in that hit songs, books, and movies are many times more popular than average, and *unpredictability*, so that well-informed experts routinely fail to identify these hits beforehand (for example, Bentley et.al. 2007, De Vany 2004, Kretschmer et.al. 1999, Walls 2005).

The very act of consumer choice in such industries is governed not just by the set of incentives described by conventional consumer demand theory, but by the choices of others (Potts et.al. 2008), so that the payoff of an individual is an explicit function of the actions of others. Schelling (1973) describes an entire set of such issues as being one of 'binary choice with externalities'.

Examination of other domains in which the events of interest are outcomes of social processes reveals a similar pattern – market crashes, regime collapses, fads and fashions, and social movements involve significant segments of society but are rarely anticipated. For example, the adoption of innovations (e.g., Arthur 1989, Rodgers 2003, Young 2005, Bettencourt et al. 2006); diffusion of criminal (e.g., Glaeser et al. 1996) and sociopolitical behaviors (e.g., Lohmann 1994, Nowak et al. 2000, Hedstrom et al. 2000, Colbaugh and Glass 2009); sales in online markets (e.g., Leskovec et al. 2006, Dhar and Chang 2007); trading in financial markets (e.g., Shiller 2000), and the rise and fall of fads and fashions (e.g., Schelling 1973, Bikhchandani et al. 1998).

In the elegant experiment described in Salganik et al. 2006, researchers constructed an online music market and examined the role social influence played in which songs participants chose to download. The experiment revealed that increasing the extent to which participants were able to observe the selections of others – that is, the strength of the social influence signal – led to an increase (decrease) in the popularity of the most (least) popular songs and a decrease in the predictability of song popularity based on quality. Other experimental studies, such as those conducted in social psychology (Asch 1953) and experimental finance reach similar conclusions regarding the effects of social influence.

The aim of this paper is to examine the extent to which the existence of social influence may, somewhat paradoxically, *increase* the extent to which winners can be identified at a very early stage in the process of consumer choices in a market (Colbaugh and Glass 2009). As noted above, in markets where social influence is important, *ex ante* prediction of eventual winners may be either very difficult or even impossible. However, once the process of choice has begun, only a very small number of purchases may be necessary to give a reasonable prospect of being able to identify the eventual winners.

Section 2 describes the data, section 3 sets out some initial analysis, and section 4 derives a prediction rule.

## 2. The data

The Salganik et.al. experiment created an artificial 'music market' in which participants downloaded previously unknown songs either with or without knowledge of previous participants' choices. Increasing the strength of social influence increased both inequality and unpredictability of success. Success was also only partly determined by quality: The best songs rarely did poorly, and the worst rarely did well, but any other result was possible.

We examined data for 18 experimental worlds, in each of which the same 48 songs were available for downloading. The detailed description of the available data for each of these worlds is described in Salganik et.al. (op.cit.), and is publicly available from the Princeton University Office of Population Research data archive: http://opr.princeton.edu/archive/. We briefly summarize the subset of this data used here. In 16 of the worlds a social signal is present. In 8 of these worlds, the person making the choice of whether or not to download was given information on the previous number of downloads carried out by other people, with the songs sorted into popularity at that time. For purposes of description, we denote these experiments as being 'strong positive externality process' or strong PEP for short

In a further 8 worlds, the same information was provided, but it was not sorted into rank order. For purposes of description, we denote these experiments as being 'weak positive externality process' or weak PEP for short Finally, in two of the worlds there is no social signal at all, designated 'no PEP'.

The total number of individual downloads in the experiments varied between 659 and 2193. Figure 1 plots the histograms of the frequency with which each song had been downloaded at the end of four of the experiments, which are entirely typical of the experiments as a whole.

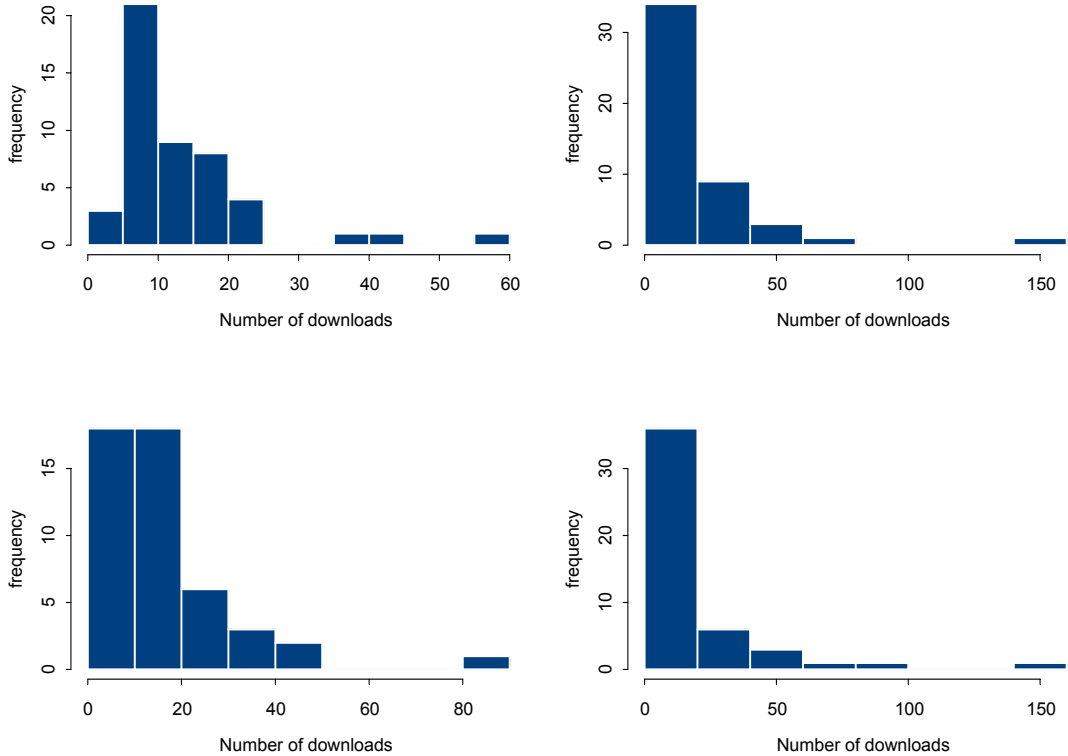

**Figure 1**  *Histograms of the frequency of downloads of the 48 individual songs at the end of four of the experiments carried out by Salganik et.al. 2006*

The right-skew nature of the outcomes is immediately clear from Figure 1. In one of these four illustrative experiments, even after 834 individual downloads from the total set of 48 songs, one song had received no downloads at all and the second lowest had just 3, compared to the highest which had 81. This ratio of highest to (non-zero) lowest of 27 reflects the inequality of the outcome, and in the other three experiments of Figure 1 this ratio was 19, 25.7 and 39.5.

Table 1 sets out in more detail information on the final outcomes in each of the experiments. Mean/median is the mean number of downloads across the 48 songs at the end of the experiment divided by the median. 'Max' is the number of downloads of the 'winner', the most frequently down loaded song, and N is the total number of downloads. Max/N is 'max' as a percentage of N. The final column is simply the identification tags we assigned to each experiment, the number themselves have no analytical significance, they are merely for identification. Experiments 11, 21, …, 81 are in fact weak PEP worlds, 12, 22, …, 82 are strong PEP worlds, and 91 and 92 no-PEP worlds.

*Table 1     Various information on the distributions of the final outcomes of the experiments*

| mean/median | max | N | max/N | experiment |
|---|---|---|---|---|
| 1.3 | 57 | 659 | 8.65 | 11 |
| 1.77 | 154 | 1021 | 15.08 | 12 |
| 1.24 | 81 | 834 | 9.71 | 21 |
| 2.02 | 158 | 968 | 16.32 | 22 |
| 1.39 | 65 | 733 | 8.87 | 31 |
| 1.96 | 114 | 892 | 12.78 | 32 |
| 1.13 | 66 | 871 | 7.58 | 41 |
| 1.7 | 165 | 1103 | 14.96 | 42 |
| 1.21 | 68 | 755 | 9.01 | 51 |
| 1.85 | 161 | 1109 | 14.52 | 52 |
| 1.16 | 61 | 944 | 6.46 | 61 |
| 1.96 | 135 | 941 | 14.35 | 62 |
| 1.17 | 69 | 1013 | 6.81 | 71 |
| 1.77 | 154 | 1149 | 13.4 | 72 |
| 1.14 | 44 | 819 | 5.37 | 81 |
| 2.27 | 179 | 926 | 19.33 | 82 |
| 1.09 | 77 | 1571 | 4.9 | 91 |
| 1 | 79 | 2193 | 3.6 | 92 |

**Notes:** *Mean/median is the mean number of downloads across the 48 songs at the end of the experiment divided by the median. 'Max' is the number of downloads of the 'winner', the most frequently down loaded song, and N is the total number of downloads. Max/N is 'max' as a percentage of N. The final column is simply the identification tags we assigned to each experiment*

### 3.     Initial evidence on predictability of outcomes

Arthur (op.cit.) gives a definition of predictability of product i after n choices by consumers as the following: denoting the market share of product i after n choices as $i_n$, it is predictable if the observer can *ex ante* construct a forecasting sequence $\{i_n^*\}$ with the property that $|i_n - i_n^*| \to 0$, with probability one, as $n \to \infty$.

Our aim in this paper is rather more heuristic. Specifically, we examine whether a rule can be discovered which will enable *ex ante* the top ranked song at the end of each experiment to be identified. In other words, we are not trying to predict the exact number of downloads (or market share) at the end of each experiment, but to see if the 'winner' of each experiment (i.e. the top ranked song at the end) can be identified *ex ante*.

Although this aim is less ambitious than that of predicting the final market share of the 'winner', it is nevertheless one which could be of considerable value in any practical situation.

The eventual winner in fact often emerges at a very early stage of each of the experiments, as the following analysis shows, of which there are two steps:

First, given two vectors x and y, the Spearman rank correlation tests the null hypothesis that the ranks of x and y are *un*correlated. For each of the experiments, we took steps 1 to n and in each case carried out the Spearman test against the data at step N, the final one in the experiment. So the first test in the sequence was the data at step 1 i.e. after the first individual download in the experiment, and the data at step N; the second the data at step 2 and the data at step N, and so on.

We identify the first step in the jth experiment– $\phi_j$ say - at which the null hypothesis is rejected at a p-value below the conventional level of significance, 0.05. So, for example, with the experiment whose final outcome is plotted at the top left of Figure 1, the correlation between the rankings at step 1 (when there is just a single observation of a song with one download which therefore has the biggest rank and all the rest are ranked equally, having no downloads) and at step 659 is 0.058, and the null hypothesis that this is zero is only rejected at p = 0.69. However, by step 10 the correlation is 0.304, and the null hypothesis that this is zero is rejected at p = 0.037.

In no fewer than 13 out of the 18 experiments, the null hypothesis is rejected at a p-value < 0.05 at one of the first 10 steps of the experiment. For the remaining 5, it is rejected at steps 25, 29, 34, 43 and 56. Recall that the experiments vary in length between 659 and 2193 steps, and it is apparent that a good approximation to the eventual rankings emerges at a very early stage.

In the second step, we then compared the rankings at step $\phi_j$ with the rankings at step $N_j$ in each case, and examined whether the eventual winner had already emerged.

In the 13 experiments where $\phi \leq 10$ on 2 occasions, there was already an unequivocal leader, which at step $N_j$ was also the 'winner'. In 3 additional experiments, there were 2 joint leaders and the eventual winner was one of these; in a further 2 there were 3 joint leaders and the eventual winner was one of these; and in a further 2 there were 4 joint leaders and the eventual winner was one of these.

In 9 out of these 13 experiments, then, the eventual winner was already unequivocally in the lead or one of a small number of the most downloaded at step $\phi_j$, $\phi_j \leq 10$.

For the 5 experiments where $\phi_j > 10$, the eventual winner was not identified at all by the winner(s) at step $\phi_j$.

So although these results are mixed, they do suggest that in half of the experiments, the eventual winner can either be identified unequivocally or as one of a group of no more than 4 out of the total of 48 songs after just 10 individual downloads have taken place. This is approximately 1 per cent of the eventual total number.

However, as a practical tool for *ex ante* prediction, these results do not constitute a rule at all, for the simple reason that the final rankings after N steps have by definition not emerged after n steps, n << N. But they do suggest that in some of the experiments, early identification of the winner may be possible.

**4.     A heuristic prediction rule**

A key characteristic of processes of agent choice or selection in which the decisions of others are taken directly into account is that the final outcome of any such process will typically exhibit considerable right-skew (for example, Simon 1955, Bentley et.al. 2009).

An obvious way in which to proceed is to examine the data on a step-by-step basis and to see at what point the outcome could be regarded as exhibiting a right-skew distribution. In other words, to examine the distribution of the number of downloads of each of the 48 songs after each individual download has taken place.

One possibility is to carry out a formal statistical test that the data follow a hypothesised right skew distribution, using the Kolmogorov-Smirnov test or, where possible, its more powerful alternative the Anderson-Darling. However, this requires that the process evolves to follow a known such distribution.

For a Gaussian distribution, the theoretical mean of the data is equal to the theoretical median. Denote by MM the ratio of the mean to the median. Of course, in an empirical setting, MM may deviate from 1 even if the data are Gaussian especially in a small sample. But the deviation is very unlikely to be large. By way of example, consider a data set of 20 observations drawn at random from a Gaussian distribution with mean 10 and standard deviation of 2 (this effectively rules out any non-zero numbers being drawn). Over 500 replications of such a sample, the empirical MM was in the range 0.914 to 1.122 and 95 per cent of the total (i.e. 475 ) were in the range 0.94 to 1.06. With a sample of 40 observations and 500 replications, the calculated MM falls in the range 0.936 to 1.066. So even with small samples, the calculated MM deviates very little from the theoretical value of 1 if the data do indeed follow a Gaussian distribution.

In contrast, in right-skew distributions, the theoretical MM is distinctly larger than 1. For an exponential distribution, with rate parameter $\lambda$, the mean is $1/\lambda$ and the median is $\log(2)/\lambda$. So the MM theoretically is $1/\log(2)$, or around 1.44. For a lognormal, where $\mu$ is the mean of the natural log of the variable and $\sigma$ is the standard deviation, the theoretical median is $\exp(\mu)$ and the theoretical mean $\exp(\mu + \sigma^2/2)$, so again MM > 1 (unless of course $\sigma^2$ is close to zero, when a lognormal is similar to a normal, though this is not the case in these experiments). And, obviously, for the power law, empirical estimates of MM will in general give a value > 1 even if the population mean does not exist.

We therefore calculated the mean/median value at each step of each experiment (though in the very early stages this ratio does not exist given than the median number of download is zero). We averaged this across the 8 'strong' and 8 'weak' positive externality experiments and across 2 experiments with no such externality.

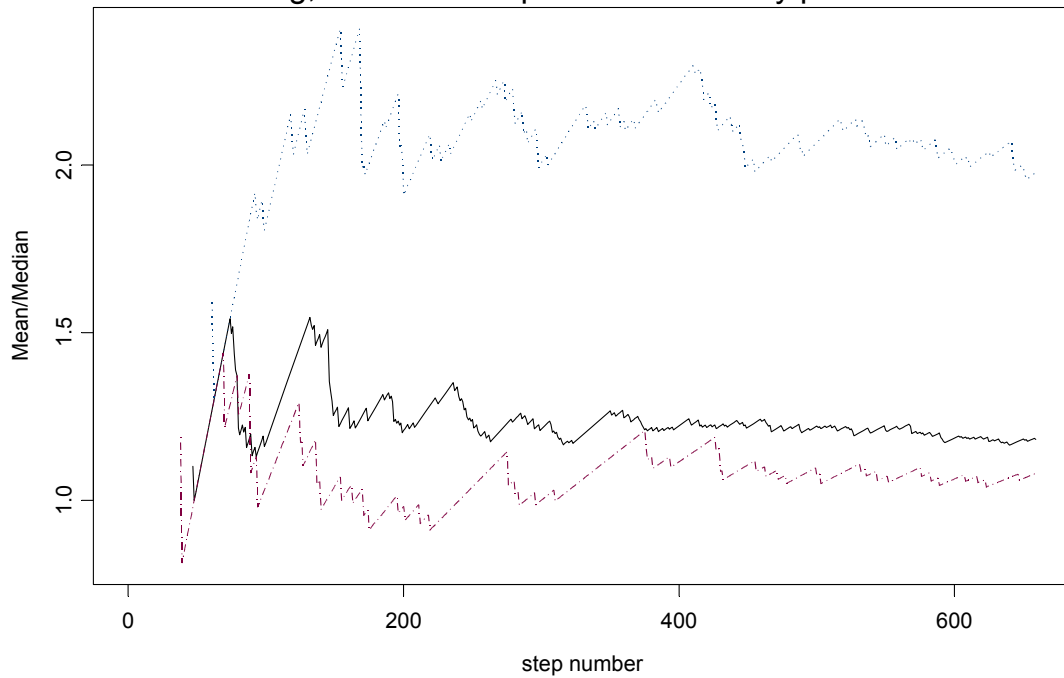

**Figure 2**     *Dotted line at top is average at step k of the mean/median of the 8 strong PEP experiments; solid line is average at step k of the mean/median of the 8 weak PEP experiments; broken line at bottom is average at step n of the mean/median for the 2 non-*

*PEP experiments.. For n close to zero, the median is zero. The data is plotted up to step 659, the length of the shortest experiment.*

It is evident that at a fairly early stage in the process, the different types of experiment become differentiated using the mean/median criterion. Note that the skew is more marked for the strong PEP than in an exponential distribution, where the theoretical mean/median is 1.44.
The next question is therefore whether the empirical mean/median is a useful tool with which to make early identification of eventual 'winners' in the experiments.

As an initial exercise, we selected the first time in each experiment that the mean/median > 1.10, with the next step also being above 1.10. We compared the rankings at this step, $\tau$ say, with the rankings in the final step, N.

Specifically, we examined whether the eventual overall winner, the one with the most downloads at time N, can be identified in any way at time $\tau$. The most obvious thing to do was to see if the single highest ranked tune at step $\tau$ was the same as the highest at time N. This was in fact the case in 4 of the experiments, all of which were strong PEPs (experiments denoted 12, 52, 72 and 82 in the table below). The next piece of information was whether one of the joint highest ranked at step $\tau$ is the eventual winner, which was the case in a further 4 experiments.

Table 2 sets out information on this, along with the percentage of total steps in the experiment which corresponds to step $\tau$.

*Table 2*     *Outcome of the use of the decision rule in identifying eventual winners*

| experiment | tau/N | maximum single download at step tau | winner at time N and winner at time tau | Joint winner at time N winner at time tau | number of joint winners |
|---|---|---|---|---|---|
| 11 | 8.04 | 2 | no | yes | 6 |
| 12 | 4.11 | 7 | yes | n/a | n/a |
| 21 | 5.4 | 9 | no | no | n/a |
| 22 | 4.75 | 3 | no | yes | 4 |
| 31 | 4.5 | 4 | no | no | n/a |
| 32 | 5.05 | 4 | no | yes | 3 |
| 41 | 3.1 | 2 | no | no | n/a |
| 42 | 3.63 | 4 | no | no | n/a |
| 51 | 7.02 | 4 | no | yes | 2 |
| 52 | 4.26 | 8 | yes | n/a | n/a |
| 61 | 3.5 | 4 | no | no | n/a |
| 62 | 5.74 | 4 | no | yes | 3 |
| 71 | 3.06 | 3 | no | yes | 2 |
| 72 | 5.31 | 13 | yes | n/a | n/a |
| 81 | 5.74 | 8 | no | no | n/a |
| 82 | 5.11 | 8 | yes | n/a | n/a |
| 91 | 2.42 | 3 | no | no | n/a |
| 92 | 1.41 | 3 | no | no | n/a |

Column 1 is simply a system for identifying the experiment in the database we used, and the numbers have no significance as such.

Column 2 shows the percentage of total steps in the experiment at which the mean/median > 1.10 for the first time. Column 3 shows the number of downloads of the market leaders at that time. Note that in general it is very small.

Column 4 indicates by yes/no whether the winner at time N at the end of the experiment was also the unequivocal leader at time τ. Column 5 indicates by yes/no whether the winner at time N was one of a group of joint leaders at time τ, and column 6 shows the number of joint leaders at time τ. So, for example, in experiment 11 at time τ, 6 songs each had 2 downloads and were the joint leaders, the rest having either 1 or 0. The eventual winner was one of this group.

In four of the experiments (12, 52, 72, and 82), the eventual winner was identified unequivocally at time τ. Step τ as a percentage of the total number of steps (individual downloads) in the experiment varied between 4.11 and 5.31, and the actual number of downloads of the winner at time τ ranged between 7 and 13.

In a further five experiments (11, 22, 51, 62 and 71), the eventual winner at step N was one of the joint winners at step τ. The number of joint winners varied between 2 and 6, and again τ was small in comparison to the total number of steps in the experiment, varying between 3.06 and 8.04.

In experiment 21, at step τ, where τ is 6.4 per cent of N, the eventual winner was placed joint second. In experiment 32, the eventual winner was third at step τ.

The rule was less successful in the other experiments, but not completely without value. So, for example, in experiment 31 at step τ, the eventual winner was one of a group of 7 which were placed joint $6^{th}$. In experiment 42, the eventual winner was one of a group of 11 which was joint $3^{rd}$. In experiment 61, the eventual winner was one of a group of 8 which was again joint $3^{rd}$.

The only experiments involving a positive externality process where the eventual winner did not emerge, either unequivocally or as part of a small group, at an early stage were experiments 41 and 81, and experiments 91 and 92 where there was no direct social influence involved.

Of the 8 experiments which exhibit strong positive externality processes, the winner at time N can always be identified very early, either unequivocally or as part of a small group, using the mean/median > 1.10 criterion. In addition, as mean/median evolves over time, it rapidly becomes apparent which experiments are strong positive externality processes.

So the simple statistic, the mean/median, appears to be a useful way of a) identifying at an early stage whether a process is governed in part by positive externalities in agent choice and b) identifying at an early stage in processes which do show evidence of positive externalities the choice which will eventually 'win' the process.

We checked the validity of the MM rule with 2 further data sets from Salganik which were not used in the process of generating the rule. These had older, more male, and more international participants that were recruited differently from those in the experiments used to develop the rule. So the two provide a useful test of the rule

In one of the data sets, the eventual winner was also the winner at time τ, when τ/N = 6.17 and the maximum number of downloads for any individual track was 10. In the other, the eventual winner was ranked second at time τ, and the eventual second was the winner at time τ. In this case, τ/N = 5.64 and the maximum number of downloads for any individual track was 9. So the rule appears to provide a reasonable heuristic for early identification of eventual winners, especially in the presence of strong social interaction.

## 5    Conclusion

In markets where social influence is important in determining whether or not an agent decides to adopt a particular mode of behavior or buy a particular product or brand, a large literature shows that successful *ex ante* prediction of the eventual winner is either very difficult or impossible.

However, the existence of social influence means that it is often possible to identify the eventual winner at a very early stage of the process of choice by participants in the market. Even if the winner cannot be identified exactly, a set from which this winner does emerge and which contains a small percentage of the total number of choices available can often be identified.

We illustrate this with an analysis of the artificial cultural market created by Salganik et.al (op.cit.). We derive a rule for early identification of the eventual winner, which we verify by using it successfully on two further experiments which were not part of the data sets used to create the rule.